\documentclass[showpacs,aps,prd]{revtex4}
\usepackage{slashed}
\input epsf

\begin{document}

\title{Fully-heavy pentaquark states}
\author{Jian-Rong Zhang}
\affiliation{Department of Physics, College of Liberal Arts and Sciences, National University of Defense Technology,
Changsha 410073, Hunan, People's Republic of China}

%\date{\today}

%%%%%%%%%%%%%%%%%%%%%%%%%%%%%%%%%%%%%%%%%%%%%%%%%%%%%%%%%%%%%%%%%%%%%
\begin{abstract}
Developing the
calculation techniques to fivefold heavy hadrons,
we perform the study of novel fully-heavy $QQQQ\bar{Q}$ pentaquark states
by the QCD sum rule approach that is firmly based on
the QCD basic theory. Numerically,
masses of fully-heavy pentaquark states
are calculated to be
$7.41^{+0.27}_{-0.31}~\mbox{GeV}$ for $cccc\bar{c}$ and
$21.60^{+0.73}_{-0.22}~\mbox{GeV}$ for $bbbb\bar{b}$, respectively. In experiment,
these predicted all-heavy pentaquark states
could be searched for in the $\Omega_{QQQ}\eta_{Q}$ invariant
mass spectrum.
\end{abstract}
\maketitle

\section{Introduction}\label{sec1}
In recent years, hunting for evidences of the multiquark
states composed of more than three quarks have attracted one's great interest.
Not long ago, LHCb Collaboration observed
a broad structure just above twice the $J/\psi$ mass
and a narrower
structure $X(6900)$
by using proton-proton collision data,
for which could possibly be all-charmed tetraquark states \cite{LHCb}.
Before this, LHCb Collaboration discovered
a narrow state $P_{c}(4312)^{+}$ in the $J/\psi p$ invariant
mass spectrum, and
separated the formerly reported
$P_{c}(4450)^{+}$ \cite{LHCb-1}
to two narrow overlapping
peaks, $P_{c}(4440)^{+}$ and $P_{c}(4457)^{+}$,
for which could be some pentaquark state candidates \cite{LHCb-2}.
On the investigations of these exotic states,
there are actually large numbers of related works and one could see
some recent reviews \cite{rev-1,rev-2,rev-3,rev-4,rev-5,rev-6,rev-7,rev-8} and references therein.
In addition, there newly appeared
a systematical study on the mass spectra of ground pentaquark states
in a modified chromo-magnetic interaction model \cite{Liu}.
Making a synthetical consideration of all these observations,
it seems very promising that pentaquark
states consist of fully-charmed quarks could be
found in the $\Omega_{ccc}\eta_{c}$ invariant
mass spectrum experimentally.

Without any light quark
contamination, fully-heavy pentaquark states are quite
ideal prototypes to refine one's knowledge on heavy quark dynamics.
To probe a fully-heavy pentaquark state, one inevitably has to
confront the very intricate nonperturbative QCD problem.
As one reliable way for calculating nonperturbative effects,
the QCD sum rule \cite{svzsum} is firmly founded on the QCD theory,
and has been successfully applied to plenty of hadronic states
\cite{overview,overview1,overview2,overview3}.
Referring to fully-heavy pentaquark states,
the related operator product expansion (OPE) calculations are quite complicated
as one has to treat many multi-loop massive propagator diagrams.
Making the development of corresponding
calculation techniques,
we devote to investigating
fully-heavy pentaquark states with the help of trustworthy
QCD sum rule method in this paper.

The rest of the paper is organized as below. The QCD sum rule
is derived for fully-heavy pentaquark states in Sec. \ref{sec2},
followed by numerical analysis and discussions in Sec.
\ref{sec3}. The last part presents a brief summary.

%%%%%%%%%%%%%%%%%%%%%%%%%%%%%%%%%%%%%%%%%%%%%%%%%%%%%%%%%%%%%%%%%%%
\section{Fully-heavy pentaquark states in QCD sum rules}\label{sec2}
Consulting interpolating currents for heavy mesons \cite{reinders} and
baryons \cite{Ioffe} in full QCD,
one can construct the following form of current
\begin{eqnarray}
j_{\mu}=(\epsilon_{abc}Q_{a}^{T}C\gamma_{\mu}Q_{b}Q_{c})(Q_{e}i\gamma_{5}\bar{Q}_{e}),
\end{eqnarray}
for $QQQQ\bar{Q}$ pentaquark states.
Here $T$ denotes matrix
transposition, $C$ means the charge conjugation matrix,
$Q$ could be the heavy charm or bottom quark, and the subscript $a$, $b$,
$c$, and $e$ are color indices.

Generally, the two-point correlator
\begin{eqnarray}
\Pi_{\mu\nu}(q^{2})=i\int
d^{4}x\mbox{e}^{iq.x}\langle0|T[j_{\mu}(x)\overline{j_{\nu}}(0)]|0\rangle,
\end{eqnarray}
can be parameterized as
\begin{eqnarray}
\Pi_{\mu\nu}(q^{2})=-g_{\mu\nu}[\slashed{q}\Pi_{1}(q^{2})+\Pi_{2}(q^{2})]+....
\end{eqnarray}
Concerning with the part proportional to $-g_{\mu\nu}\slashed{q}$,
matching its descriptions at the hadron level
and at the quark level, and applying a Borel transform,
one arrives at
\begin{eqnarray}\label{sumrule}
\lambda^{2}e^{-M_{H}^{2}/M^{2}}&=&\int_{25m_{Q}^{2}}^{s_{0}}ds\rho e^{-s/M^{2}},
\end{eqnarray}
in which $M_{H}$ is the studied hadron's mass and the spectral density
$\rho=\frac{1}{\pi}\mbox{Im}\Pi_{1}(s)$.
Taking
the derivative of Eq. (\ref{sumrule}) with respect to $-\frac{1}{M^2}$ and then dividing
the result by Eq. (\ref{sumrule}) itself, one could acquire the mass
\begin{eqnarray}\label{sum rule}
M_{H}&=&\sqrt{\int_{25m_{Q}^{2}}^{s_{0}}ds\rho s
e^{-s/M^{2}}/
\int_{25m_{Q}^{2}}^{s_{0}}ds\rho e^{-s/M^{2}}}.
\end{eqnarray}

In the OPE calculation, one works
at the momentum-space with the heavy-quark propagator
\cite{reinders}, and
then the result is dimensionally
regularized at $D=4$, by extending the interrelated techniques \cite{Nielsen,Zhang}
to fully-heavy pentaquark systems.
Concretely, the spectral density $\rho=\rho^{\mbox{pert}}+\rho^{\langle
g^{2}G^{2}\rangle}+\rho^{\langle
g^{3}G^{3}\rangle}$ is expressed as
\begin{eqnarray}
\rho^{\mbox{pert}}&=&\frac{3}{5\cdot2^{14}\pi^{8}}\int_{\alpha_{min}}^{\alpha_{max}}\frac{d\alpha}{\alpha^{3}}\int_{\beta_{min}}^{\beta_{max}}\frac{d\beta}{\beta^{3}}\int_{\gamma_{min}}^{\gamma_{max}}\frac{d\gamma}{\gamma^{3}}
\int_{\xi_{min}}^{\xi_{max}}\frac{d\xi}{\xi^{3}}\frac{\textbf{h}^{3}(m_{Q}^{2}-\textbf{h}s)^{3}}{(1-\alpha-\beta-\gamma-\xi)^{3}}
\bigg[-8\textbf{h}^{2}(m_{Q}^{2}-\textbf{h}s)^{2}\nonumber\\
&+&(35\textbf{h}^{3}s+25\alpha\beta\textbf{h}m_{Q}^{2})(m_{Q}^{2}-\textbf{h}s)-20\textbf{h}^{4}s^{2}-40\alpha\beta\textbf{h}^{2}m_{Q}^{2}s-20\alpha\beta\gamma\xi m_{Q}^{4}\bigg],\nonumber
\end{eqnarray}

\begin{eqnarray}
\rho^{\langle g^{2}G^{2}\rangle}&=&\frac{m_{Q}^{2}\langle g^{2}G^{2}\rangle}{2^{14}\pi^{8}}\int_{\alpha_{min}}^{\alpha_{max}}\frac{d\alpha}{\alpha^{3}}\int_{\beta_{min}}^{\beta_{max}}\frac{d\beta}{\beta^{3}}\int_{\gamma_{min}}^{\gamma_{max}}\frac{d\gamma}{\gamma^{3}}
\int_{\xi_{min}}^{\xi_{max}}\frac{d\xi}{\xi^{3}}\frac{\textbf{h}^{3}}{(1-\alpha-\beta-\gamma-\xi)^{3}}\bigg\{\Big[(1-\alpha-\beta\nonumber\\
&-&\gamma-\xi)^{3}
+2\beta^{3}+2\xi^{3}\Big]\Big[-4\textbf{h}^{2}(m_{Q}^{2}-\textbf{h}s)^{2}+7\textbf{h}^{3}s(m_{Q}^{2}-\textbf{h}s)-\textbf{h}^{4}s^{2}
-\alpha\beta\gamma\xi m_{Q}^{4}+3\alpha\beta\textbf{h}m_{Q}^{4}\nonumber\\
&-&4\alpha\beta\textbf{h}^{2}m_{Q}^{2}s\Big]+\alpha\beta(1-\alpha-\beta-\gamma-\xi)^{3}\textbf{h}m_{Q}^{2}(2m_{Q}^{2}-3\textbf{h}s)
+2\gamma\xi(\beta^{3}+\xi^{3})\textbf{h}m_{Q}^{2}(2m_{Q}^{2}-3\textbf{h}s)\nonumber\\
&-&6\alpha\beta\gamma\xi(\beta^{2}+\xi^{2})m_{Q}^{2}(m_{Q}^{2}-\textbf{h}s)-6(\gamma\xi^{3}+\alpha\beta^{3})\textbf{h}^{2}s(m_{Q}^{2}-\textbf{h}s)
+3\textbf{h}(3\alpha\beta^{3}+2\gamma\xi^{3})(m_{Q}^{2}-\textbf{h}s)^{2}
\bigg\},\nonumber
\end{eqnarray}

and
\begin{eqnarray}
\rho^{\langle g^{3}G^{3}\rangle}&=&\frac{\langle g^{3}G^{3}\rangle}{2^{16}\pi^{8}}\int_{\alpha_{min}}^{\alpha_{max}}\frac{d\alpha}{\alpha^{3}}\int_{\beta_{min}}^{\beta_{max}}\frac{d\beta}{\beta^{3}}\int_{\gamma_{min}}^{\gamma_{max}}\frac{d\gamma}{\gamma^{3}}
\int_{\xi_{min}}^{\xi_{max}}\frac{d\xi}{\xi^{3}}
\frac{\textbf{h}^{3}}{(1-\alpha-\beta-\gamma-\xi)^{3}}
\bigg\{\Big[(1-\alpha-\beta-\gamma-\xi)^{3}\nonumber\\
&+&2\beta^{3}+2\xi^{3}\Big]\Big[-4\textbf{h}^{2}(m_{Q}^{2}-\textbf{h}s)^{2}+7\textbf{h}^{3}s(m_{Q}^{2}-\textbf{h}s)-\textbf{h}^{4}s^{2}\Big]
+2\alpha\beta\Big[(1-\alpha-\beta-\gamma-\xi)^{3}
+6\beta^{3}+\xi^{3}\Big]\nonumber\\
&\times&\textbf{h}m_{Q}^{2}(3m_{Q}^{2}-4\textbf{h}s)-\alpha\beta\gamma\xi\Big[(1-\alpha-\beta-\gamma-\xi)^{3}
+12\beta^{3}+12\xi^{3}
\Big]m_{Q}^{4}-\alpha\beta(1-\alpha-\beta-\gamma-\xi)^{3}\nonumber\\
&\times&\textbf{h}m_{Q}^{2}(m_{Q}^{2}-\textbf{h}s)+2\gamma\xi(\beta^{3}+6\xi^{3})\textbf{h}m_{Q}^{2}(2m_{Q}^{2}-3\textbf{h}s)
-2\Big[(1-\alpha-\beta-\gamma-\xi)^{4}+2\beta^{4}+2\xi^{4}\Big]\nonumber\\
&\times&\textbf{h}^{2}m_{Q}^{2}(8m_{Q}^{2}-15\textbf{h}s)+2\textbf{h}m_{Q}^{4}\Big[5\alpha\beta(1-\alpha-\beta-\gamma-\xi)^{4}+2(3\alpha\beta+2\gamma\xi)(\beta^{4}+\xi^{4})\Big]
\bigg\}.\nonumber
\end{eqnarray}

It is
defined as $\textbf{h}=\frac{1}{\frac{1}{\alpha}+\frac{1}{\beta}+\frac{1}{\gamma}+\frac{1}{\xi}+\frac{1}{1-\alpha-\beta-\gamma-\xi}}$,
and the integration limits of $\alpha$, $\beta$, $\gamma$, and $\xi$ are given by
\begin{eqnarray}
\alpha&=&\frac{1}{2}\Bigg[\bigg(1-\frac{15m_{Q}^{2}}{s}\bigg)\pm\sqrt{\bigg(1-\frac{15m_{Q}^{2}}{s}\bigg)^{2}-\frac{4m_{Q}^{2}}{s}}\Bigg], \nonumber\\
\beta&=&\frac{1}{2}\Bigg[\Big(1-\alpha-\frac{8\alpha m_{Q}^{2}}{\alpha s-m_{Q}^{2}}\Big)\pm\sqrt{\Big(1-\alpha-\frac{8\alpha m_{Q}^{2}}{\alpha s-m_{Q}^{2}}\Big)^{2}-\frac{4\alpha(1-\alpha)m_{Q}^{2}}{\alpha s-m_{Q}^{2}}}\Bigg],\nonumber\\
\gamma&=&\frac{1}{2}\Bigg[\Big(1-\alpha-\beta-\frac{3}{\frac{s}{m_{Q}^{2}}-\frac{1}{\alpha}-\frac{1}{\beta}}\Big)\pm\sqrt{\Big(1-\alpha-\beta-\frac{3}{\frac{s}{m_{Q}^{2}}-\frac{1}{\alpha}-\frac{1}{\beta}}\Big)^{2}-4\frac{1-\alpha-\beta}{\frac{s}{m_{Q}^{2}}-\frac{1}{\alpha}-\frac{1}{\beta}}}\Bigg],\nonumber
\end{eqnarray}
and
\begin{eqnarray}
\xi&=&\frac{1}{2}\Bigg[\Big(1-\alpha-\beta-\gamma\Big)\pm\sqrt{\Big(1-\alpha-\beta-\gamma\Big)^{2}-4\frac{1-\alpha-\beta-\gamma}{\frac{s}{m_{Q}^{2}}-\frac{1}{\alpha}-\frac{1}{\beta}-\frac{1}{\gamma}}}\Bigg].\nonumber
\end{eqnarray}

As the usual
QCD sum rule treatment, it mainly takes into account the well-known two-gluon
condensate $\langle g^{2}G^{2}\rangle$ and three-gluon condensate $\langle g^{3}G^{3}\rangle$ here.
Note that it was discussed that including higher-dimension gluonic condensates
may be helpful to the sum rule analysis to some extent.
In that case, those higher-dimension gluonic
operators would require some speculative input assumptions for their numerical values.
Generally speaking, the higher the dimension of the gluonic operator,
the less known the value of its matrix element.
By way of parenthesis, involving those unascertained
higher dimensional gluonic condensates, one technically has to accomplish
a tremendous task since a number of
multi-loop massive propagator diagrams need to be treated for the
present fully-heavy pentaquark states, for
which could be taken into account in some further work.

\section{Numerical analysis and discussions}\label{sec3}
The input parameters are taken as
$\langle g^{2}G^{2}\rangle=0.88\pm0.25~\mbox{GeV}^{4}$ and $\langle
g^{3}G^{3}\rangle=0.58\pm0.18~\mbox{GeV}^{6}$ \cite{svzsum,overview3,Narison},
and $m_{Q}$ is set as the running charm mass
$m_{c}=1.27\pm0.02~\mbox{GeV}$ \cite{PDG} at first.
As one knows, the QCD sum rule method has made approximations
in the OPE of the correlation functions and
introduced a very complicated and largely unknown
structure of the hadronic dispersion integrals in the phenomenological side.
In this way, complying with the
criterion of sum rule analysis, one should find
appropriate work windows for both the continuum threshold $\sqrt{s_{0}}$ and the Borel
parameter $M^{2}$ in
which the two sides of QCD sum rules have a good overlap and information on the hadronic resonance can be reliably extracted.
In phenomenology, the threshold $\sqrt{s_{0}}$ is the energy which
characterizes the beginning of the continuum state
and the gap $\sqrt{s_{0}}-M_{H}$
is typically evaluated to be about $0.3\sim0.8~\mbox{GeV}$ \cite{overview3}.
Meanwhile, the proper Borel window of $M^{2}$ can be found by analyzing the OPE convergence and the pole dominance:
the lower value of $M^{2}$ is obtained by considering the OPE
convergence, and
the upper one of $M^{2}$ is get from
the condition that the pole contribution
should be bigger than the continuum contribution.

At the start, the input parameters are kept at their central values.
The OPE convergence for the $cccc\bar{c}$ pentaquark state
can be analyzed by comparing the relative contributions of various
condensates from sum rule (\ref{sumrule}).
It is noted that the three-gluon
condensate $\langle g^{3}G^{3}\rangle$ with dimension six is much smaller than
two-gluon condensate $\langle
g^{2}G^{2}\rangle$ or perturbative contribution.
In view of the OPE convergence analysis,
the lower bound of $M^{2}$ is numerically taken as $M^{2}\geq3.5~\mbox{GeV}^{2}$.
Phenomenologically, one could fix the upper bound of $M^{2}$ according to
the pole dominance requirement. For example,
by making the comparison
between pole and continuum contribution from sum rule (\ref{sumrule})
for $\sqrt{s_{0}}=8.0~\mbox{GeV}$ in FIG. 1, one notes that
the relative pole contribution
is around $50\%$ at $M^{2}=4.8~\mbox{GeV}^{2}$ and
it descends with $M^{2}$.
Accordingly, the pole dominance condition could be satisfied
when $M^{2}\leq4.8~\mbox{GeV}^{2}$, and the Borel window is chosen as
$M^{2}=3.5\sim4.8~\mbox{GeV}^{2}$ for $\sqrt{s_0}=8.0~\mbox{GeV}$.
With an eye to the typical gap
between the continuum and resonance, the variation of threshold $\sqrt{s_{0}}$
is taken as $\sqrt{s_{0}}=7.7\sim8.2~\mbox{GeV}$ for $cccc\bar{c}$.
By way of the similar analysis as above, the corresponding Borel
windows are fixed as $M^{2}=3.5\sim4.1~\mbox{GeV}^{2}$
for $\sqrt{s_0}=7.7~\mbox{GeV}$, and
$M^{2}=3.5\sim5.3~\mbox{GeV}^{2}$ for $\sqrt{s_0}=8.2~\mbox{GeV}$, respectively.

\begin{figure}[htb!]
\centerline{\epsfysize=7.50truecm\epsfbox{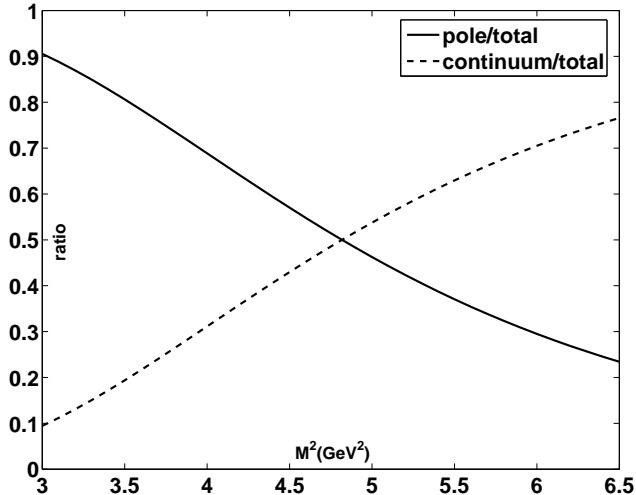}}
\caption{The phenomenological contribution in sum rule
(\ref{sumrule}) for $\sqrt{s_{0}}=8.0~\mbox{GeV}$ for
the fully-heavy $cccc\bar{c}$ pentaquark state.
The solid line is the relative pole contribution
as a function of $M^2$ and the dashed line is the relative continuum
contribution.}
\end{figure}

The Borel curves for the $cccc\bar{c}$ pentaquark state are shown in FIG. 2,
and in the chosen work windows its mass is extracted to be
$7.41^{+0.20}_{-0.23}~\mbox{GeV}$.
After varying the input parameters, the achieved mass is
$7.41^{+0.20+0.07}_{-0.23-0.08}~\mbox{GeV}$ (the first uncertainty from
the variation of threshold $\sqrt{s_{0}}$ and the Borel
parameter $M^{2}$, and the second one due to the uncertainty of QCD parameters)
or compactly $7.41^{+0.27}_{-0.31}~\mbox{GeV}$.

\begin{figure}[htb!]
\centerline{\epsfysize=7.50truecm
\epsfbox{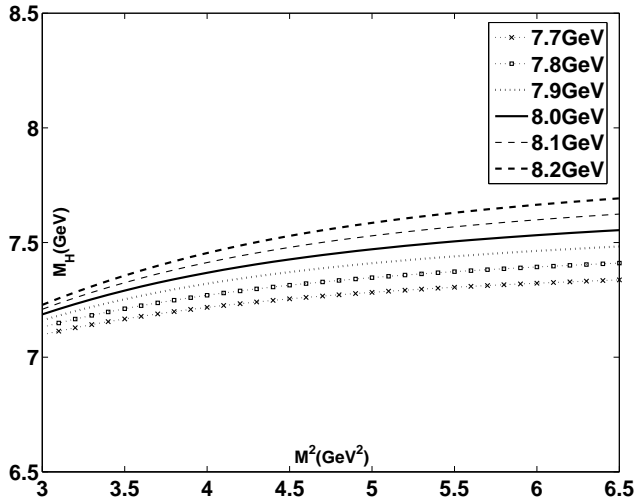}}\caption{The mass $M_{H}$
dependence on $M^2$ for the
fully-heavy $cccc\bar{c}$ pentaquark state
 from sum rule (\ref{sum rule}) is shown.
The Borel windows of $M^{2}$ are $3.5\sim4.1~\mbox{GeV}^{2}$ for
$\sqrt{s_0}=7.7~\mbox{GeV}$, $3.5\sim4.8~\mbox{GeV}^{2}$
for $\sqrt{s_0}=8.0~\mbox{GeV}$, and
$3.5\sim5.3~\mbox{GeV}^{2}$ for $\sqrt{s_0}=8.2~\mbox{GeV}$, respectively.}
\end{figure}

Replacing the heavy $m_{Q}$ by the running bottom mass
$m_{b}=4.18_{-0.02}^{+0.03}~\mbox{GeV}$ \cite{PDG},
one could straightway put forward the corresponding analysis
for fully-bottomed $bbbb\bar{b}$ pentaquark state.
Considering the much bigger mass of $bbbb\bar{b}$,
the typical gap may be somewhat small for
the fully-heavy $bbbb\bar{b}$ pentaquark state.
Thus, the gap has been further broadened and
the associated theoretical uncertainty has been included
to give a conservative mass prediction for $bbbb\bar{b}$.
Correspondingly, the threshold $\sqrt{s_{0}}$ for $bbbb\bar{b}$
is varied as $21.8\sim23.0~\mbox{GeV}$, and
its Borel curves are
displayed in FIG. 3.
Having taken into the uncertainty of both work windows and
QCD parameters, the mass value
is computed to be $21.60^{+0.73}_{-0.22}~\mbox{GeV}$
for the $bbbb\bar{b}$ pentaquark state.

\begin{figure}[htb!]
\centerline{\epsfysize=7.50truecm
\epsfbox{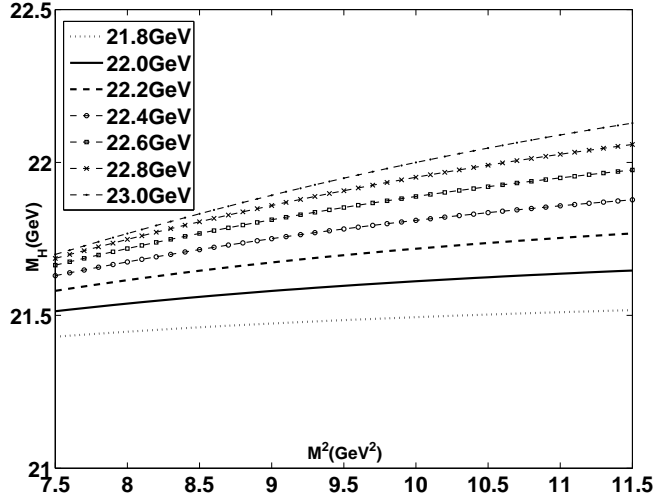}}\caption{The mass $M_{H}$
dependence on $M^2$ for the
fully-heavy $bbbb\bar{b}$ pentaquark state
 from sum rule (\ref{sum rule}) is shown.
The Borel windows of $M^{2}$ are $8.0\sim8.4~\mbox{GeV}^{2}$ for
$\sqrt{s_0}=21.8~\mbox{GeV}$, $8.0\sim9.6~\mbox{GeV}^{2}$
for $\sqrt{s_0}=22.0~\mbox{GeV}$,
$8.0\sim11.0~\mbox{GeV}^{2}$ for $\sqrt{s_0}=22.2~\mbox{GeV}$,
and $8.0\sim11.5~\mbox{GeV}^{2}$ for $\sqrt{s_0}=23.0~\mbox{GeV}$, respectively.}
\end{figure}

%%%%%%%%%%%%%%%%%%%%%%%%%%%%%%%%%%%%%%%%%%%%%%%%%%%%%%%%%%%%%%%%%%%
\section{Summary}\label{sec4}
By making the development of
calculation techniques to fivefold heavy pentaquark states,
we present the investigation of fully-heavy $QQQQ\bar{Q}$
pentaquark states from trustable QCD sum rules.
Eventually, their mass spectrums
are predicted to be $7.41^{+0.27}_{-0.31}~\mbox{GeV}$
for the $cccc\bar{c}$ state, and
$21.60^{+0.73}_{-0.22}~\mbox{GeV}$ for the $bbbb\bar{b}$ state, respectively.
It is proposed that these states could be experimentally looked for in the $\Omega_{QQQ}\eta_{Q}$ invariant
mass spectrum.
For the future, one can expect that further theoretical studies
and experimental efforts
may shed more light on the nature of exotic
fully-heavy pentaquark states.

%%%%%%%%%%%%%%%%%%%%%%%%%%%%%%%%%%%%%%
\begin{acknowledgments}
This work was supported by the National
Natural Science Foundation of China under Contract
Nos. 11475258 and 11675263, and by the project for excellent youth talents in
NUDT.
\end{acknowledgments}

%%%%%%%%%%%%%%%%%%%%%%%%%%%%%%%%%%%%%%%%%%%%%%%%%%%%


\begin{thebibliography}{99}
\bibitem{LHCb}R.~Aaij {\it et al.} (LHCb Collaboration), Sci. Bull. {\bf65}, 1983 (2020).


\bibitem{LHCb-1}R.~Aaij {\it et al.} (LHCb Collaboration), Phys. Rev. Lett.
{\bf115}, 072001 (2015).


\bibitem{LHCb-2}R.~Aaij {\it et al.} (LHCb Collaboration), Phys. Rev. Lett. {\bf122}, 222001
(2019).


\bibitem{rev-1}H.~X.~Chen, W.~Chen, X.~Liu, and S.~L.~Zhu,
Phys. Rept. {\bf639}, 1 (2016).


\bibitem{rev-2}A.~Esposito, A.~Pilloni, and A.~D.~Polosa, Phys. Rept. {\bf668}, 1  (2016).


\bibitem{rev-3}A.~Ali, J.~S.~Lange, and S.~Stone, Prog. Part. Nucl. Phys. {\bf97},
123 (2017).

\bibitem{rev-4}R.~F.~Lebed, R.~E.~Mitchell, and E.~S.~Swanson,  Prog. Part. Nucl. Phys.
{\bf93}, 143 (2017).

\bibitem{rev-5}F.~K.~Guo, C.~Hanhart, U.~G.~Meissner, Q.~Wang, Q.~Zhao, and B.~S.~Zou, Rev.
Mod. Phys. {\bf90}, 015004 (2018).



\bibitem{rev-6}Y.~R.~Liu, H.~X.~Chen, W.~Chen, X.~Liu, and S. L. Zhu,
Prog. Part. Nucl. Phys. {\bf107}, 237 (2019).

\bibitem{rev-7}N.~Brambilla, S.~Eidelman, C.~Hanhart, A.~Nefediev,
C.~P.~Shen, C.~E.~Thomas, A.~Vairo, and C.~Z.~Yuan,
Phys. Rept. {\bf873}, 1 (2020).



\bibitem{rev-8}G.~Yang, J.~L.~Ping, and J.~Segovia, arXiv:2009.00238 [hep-ph].


\bibitem{Liu}H.~T.~An, K.~Chen, and X.~Liu, arXiv:2010.05014 [hep-ph].


\bibitem{svzsum}M.~A.~Shifman, A.~I.~Vainshtein, and V.~I.~Zakharov, Nucl. Phys. {\bf B147}, 385 (1979); {\bf B147}, 448 (1979);
 V.~A.~Novikov, M.~A.~Shifman, A.~I.~Vainshtein, and V.~I.~Zakharov, Fortschr. Phys. {\bf 32}, 585 (1984).



\bibitem{overview}B.~L.~Ioffe, in \emph{The Spin Structure of The Nucleon}, edited by
B.~Frois, V.~W.~Hughes, and N.~de Groot (World Scientific,
Singapore, 1997).



\bibitem{overview1}S.~Narison, Camb. Monogr. Part. Phys. Nucl. Phys. Cosmol. {\bf17}, 1
(2002), arXiv:hep-ph/0205006.




\bibitem{overview2}P.~Colangelo and A.~Khodjamirian, in \emph{At the Frontier of
Particle Physics: Handbook of QCD}, edited by M.~Shifman,
Boris Ioffe Festschrift Vol. 3 (World Scientific,
Singapore, 2001), pp. 1495-1576.



\bibitem{overview3}M.~Nielsen, F.~S.~Navarra, and S.~H.~Lee, Phys. Rep. {\bf497},
41 (2010).



\bibitem{reinders}L.~J.~Reinders, H.~R.~Rubinstein, and S.~Yazaki, Phys. Rep. {\bf 127}, 1 (1985).


\bibitem{Ioffe}B.~L.~Ioffe, Nucl. Phys. {\bf B188}, 317 (1981);
E.~V.~Shuryak, Nucl. Phys. {\bf B198}, 83 (1982).




\bibitem{Nielsen}H.~Kim and Y.~Oh, Phys. Rev. D {\bf72}, 074012 (2005);
M.~E.~Bracco, A.~Lozea, R.~D.~Matheus, F.~S.~Navarra, and
M.~Nielsen, Phys. Lett. B {\bf624}, 217 (2005); R.~D.~Matheus,
S.~Narison, M.~Nielsen, and J.~M.~Richard, Phys. Rev. D {\bf75},
014005 (2007).



\bibitem{Zhang}J.~R.~Zhang and M.~Q.~Huang, Phys. Lett. B {\bf 674}, 28 (2009);
J.~R.~Zhang, Phys. Rev. D {\bf87}, 076008 (2013); Phys. Rev. D {\bf89}, 096006 (2014);
Phys. Rev. D {\bf103}, 014018 (2021).



\bibitem{Narison} S.~Narison, Phys. Rep. {\bf84}, 263 (1982);
G.~Launer, S.~Narison, and R.~Tarrach, Z. Phys. C {\bf26}, 433 (1984);
S.~Narison, Phys. Lett. B {\bf673}, 30 (2009).


\bibitem{PDG}M.~Tanabashi {\it et al.} (Particle Data Group), Phys. Rev. D {\bf98}, 030001 (2018) and 2019 update.




\end{thebibliography}
\end{document}